# BEAMED RADIATION IN RADIO GALAXIES

G.V. BICKNELL
*Mt. Stromlo & Siding Spring Observatories, Australian National University*

**ABSTRACT** This paper focuses on two aspects of the Physics of Radio Galaxies in which it has been proposed that relativistic beaming of radiation is necessary to account for the observations. Physical arguments are presented which show that it is quite plausible that the jets in FRI radio galaxies are relativistic on the parsec scale and decelerate to subrelativistic flow on the kiloparsec scale. On the other hand it is shown that beaming of the optical continuum from the core of Centaurus A is not necessary to account for the excitation of the emission line filaments in the ISM of this galaxy. Models involving autoionizing shocks produced by the interaction between a low Mach number jet and dense clouds along its path are capable of explaining the morphology of the filaments and the simultaneous high excitation and low excitation lines in their spectra.

## INTRODUCTION

The difference between FRI and FRII radio jets is often expressed in terms of turbulent, transonic and subrelativistic, for the former and supersonic and possibly relativistic for the latter. The kpc-scale emission from FRI's is thought to be largely unbeamed; beaming is widely thought to be implicated in the one-sided jet structures of class II radio sources and quasars. However, this standard dichotomy is not without some outstanding issues. For example, (1) The kpc-scale bases of FRI jets are frequently one-sided; (2) The small number of FRI jets that have been observed with VLBI on the pc-scale show a one-sided jet pointing to the brighter kpc-scale jet; (3) BL Lac objects, in which beaming and other relativistic effect are obviously important are believed to originate from a parent population of FRI radio galaxies, implying that the jets in the latter are relativistic on the parsec scale. (See, for example Ulrich, 1989. ) (4) If FRI jets are relativistic on the parsec scale, why are the motions of knots in at least some pc-scale FRI jets, *sub*luminal (Reid *et al.*, 1989; Venturi *et al.*, 1993). An obvious solution to some of these problems is for FRI jets to decelerate from a relativistic to a subrelativistic velocity somewhere between the parsec and kiloparsec scales. Since deceleration is implicated in the physics of the kiloparsec scale structure anyway (e.g. Bicknell, 1986) this does not seem too radical a suggestion. Nevertheless, it remains to be shown that this is dynamically consistent.

Optical observations of emission line regions in radio galaxies present another series of challenges, for the resolution of which a number of workers have also sought an explanation involving the physics of beamed radiation. For ex-ample, there is a well known deficit between the ionizing flux detected from the core of a radio galaxy and the flux required to account for the extended Balmer line luminosity. Moreover, there is a serious problem in that the temperature of Extended Emission Line Regions (inferred from the [OIII]$\lambda$4363/$\lambda$5007 ratio) is higher than predicted by standard power-law photoionization models. In some cases the "beaming" may be induced by obscuration in a thick torus surrounding the active core; in others the beaming must be relativistic to be consistent with the lack of reradiated IR flux from a putative obscuring torus. In this paper I summarize a different approach to the emission line data on one important source Centaurus A. This summary is based on collaborative work with Ralph Sutherland, Mike Dopita and David Singleton. Our approach involves the physics of autoionizing shocks and the theory has ramifications for line-emitting gas in numerous other active galaxy contexts.

## RELATIVISTIC ENTRAINING FLOW

Since there is no known expression for turbulent viscosity in the case of non-relativistic supersonic flow, let alone relativistic flow, an approach based on jet conservation laws (Bicknell, 1993) is useful. Such laws are obtained by integrating the relativistic hydrodynamic equations throughout a control volume incorporating the jet and the surrounding ISM. These laws relate jet parameters at an arbitrary cross-section (area $A_2$) to core-jet parameters (cross-sectional area $A_1$). Neglecting the initial rest-mass flux of the pc-scale jet, particle number, momentum and energy conservation imply that

$$\gamma_2 \, \rho_2 \, v_2 \, A_2 \approx -\int_{S_E} \rho \, \mathbf{v} \cdot \mathbf{n} \, dS \tag{1}$$

$$\begin{aligned}(\gamma_2^2 \frac{w}{c^2} v_2^2 + \Delta p_2) \, A_2 &- \left(\gamma_1^2 \frac{w}{c^2} v_1^2 + \Delta p_1\right) A_1 \\ &\approx -\int_{r_1}^{r_2} dr \left[\frac{dp_{\rm ext}}{dr} \int_A \left(1 - \frac{\rho^*}{\rho_{\rm ext}}\right) \frac{z}{r} \, dS\right]\end{aligned} \tag{2}$$

and

$$\left[(\gamma_2^2 - \gamma_2)\rho_2 \, c^2 + 4\,\gamma_2^2 \, p_2\right] \, v_2 \, A_2 - 4\,\gamma_1^2 \, p_1 \, v_1 \, A_1 \approx -\int_{S_E} \rho \, h \, \mathbf{v}_{\rm ent} \cdot \mathbf{n} \, dS \tag{3}$$

Here, $S_E$ is the "entrainment surface" located in the ISM.

The jet momentum is affected by the external pressure gradient. However, if the jet is initially free or of high Mach number, momentum is approximately conserved. The jet energy is affected by the entrainment of internal energy which, for the high enthalpy jets under consideration, is negligible. With these simplifications, the above equations become:

$$\gamma_2^2 \, \beta_2^2 \, (1 + \mathcal{R}) \approx \frac{1 + 4\gamma_1^2\beta_1^2}{4} \left(\frac{p_1 \, A_1}{p_2 \, A_2}\right) \tag{4}$$

$$[(\gamma_2 - 1)\,\mathcal{R} + \gamma_2] \; \gamma_2 \, \beta_2 \approx \gamma_1^2 \, \beta_1 \left(\frac{p_1 \, A_1}{p_2 \, A_2}\right) \tag{5}$$

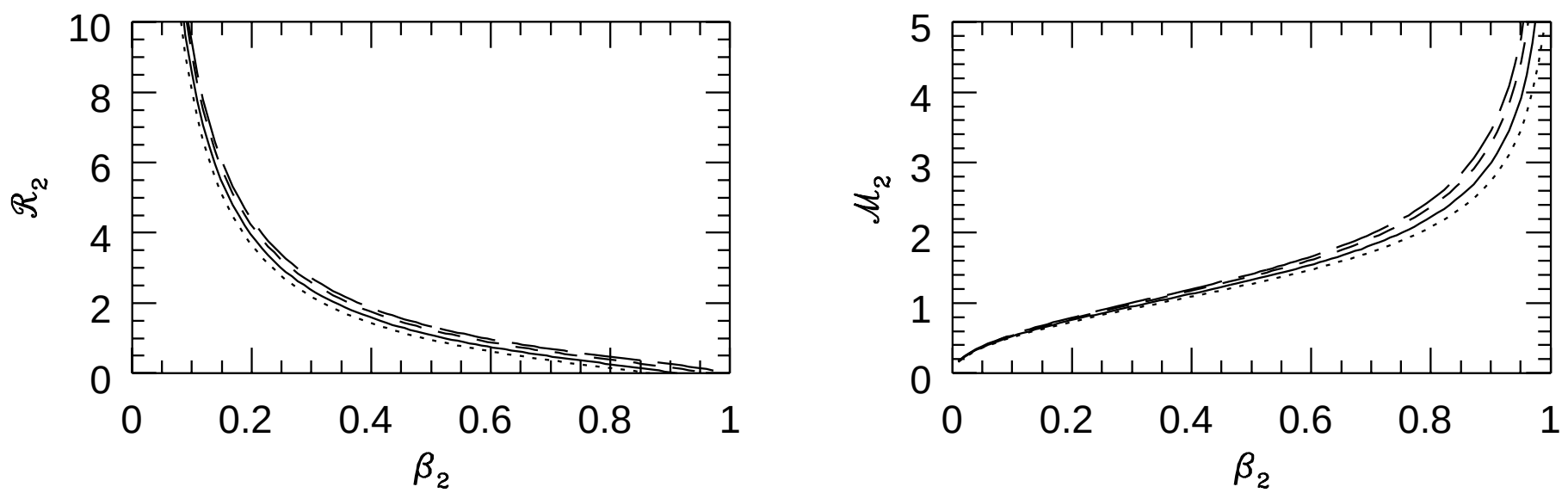

FIGURE I The ratio R of rest-mass energy density to enthalpy (left panel) and Mach number as a function of jet velocity (right panel) in an initially ultrarelativistic entraining jet. Solid curve: $\gamma_1 = 5$; dotted curve: $\gamma_1 = 2.5$; long-dashed curve: $\gamma_1 = 1.25$; short-dashed curve: $\gamma_1 = 1.1$

where $\mathcal{R} = \rho\, c^2/4p$, the ratio of jet rest-mass energy to enthalpy, and $\beta = v/c$. (The jet Mach number is given by $\mathcal{M}^2 = (2 + 3\mathcal{R})\gamma^2\beta^2$). Given jet pressures and cross-sections on both parsec and kiloparsec scales then one may estimate both $\mathcal{R}$ and $\beta$ as a function of the initial jet Lorentz factor. Moreover, $\mathcal{R}$ can be determined as a function of jet velocity and initial $\beta$ from the above equations. The parameter $\mathcal{R}$ and Mach number are plotted in figure I for a number of initial Lorentz factors. These solutions are extremely interesting since they show that when an initially ultra-relativistic jet slows to a Mach number say between 1 and 2, then its velocity is necessarily between 0.3c and 0.7c. Since the spreading rates of FRI jets are most naturally associated with transonic turbulence then this defines a natural velocity $\sim 0.3 - 0.7c$ for the kpc-scale bases of FRI sources. Such velocities are sufficient to explain the surface brightness ratios at the bases of Class I jets (see the papers by Parma and Laing in these proceedings). A velocity $\sim 0.3c$ corresponding to a Mach number of 1 has a simple physical interpretation: The ratio of energy to momentum fluxes in an ultrarelativistic jet is $c$. When the jet has slowed to a mildly relativistic velocity the same (conserved) ratio is $3v(1 + \mathcal{M}^2/6)\mathcal{M}^{-2}$. Equating the two gives, for $\mathcal{M} = 1$, $v/c = 2/7 \approx 0.29$.

## DETAILED CONSIDERATION OF NGC 315

Figure II shows the calculated Mach number of the NGC 315 jet (Willis *et al.*, 1981; Venturi *et al.*, 1993) at $\Theta = 17''$ and $27''$ as a function of the initial Lorentz factor for $p/p_{\rm min} = 10$ and 100 and for jet inclinations of 90° and 30°. The implication of these plots is that a kpc-scale transonic Mach number and a moderate initial Lorentz factor in the 90° jet are consistent if the ratio of initial jet pressure to the minimum value is not much more than about 10. If the jet pressure is much higher than this then initially, the jet is at most mildly relativistic. The allowable range of pc-scale Lorentz factors increases if the jet is inclined at 30° to the line of sight (although for $p/p_{\rm min} \approx 100$ there is not much difference in the two curves because the jet does not decelerate for such a high initial pressure). The higher Lorentz factors at lower pressures are realted to the lower inferred minimum pressure, necessitating a higher velocity for given energy and momentum fluxes. The energy flux is also of interest since it is conserved and in principle, can be estimated from minimum energy and age estimates of the lobes. A roundabout estimate for NGC 315 is $\sim 10^{43}$ergs s$^{-1}$. This is more easily achieved for a range of $p/p_{\rm min}$ if the jet is inclined at 30° to the line of sight (see figure II).

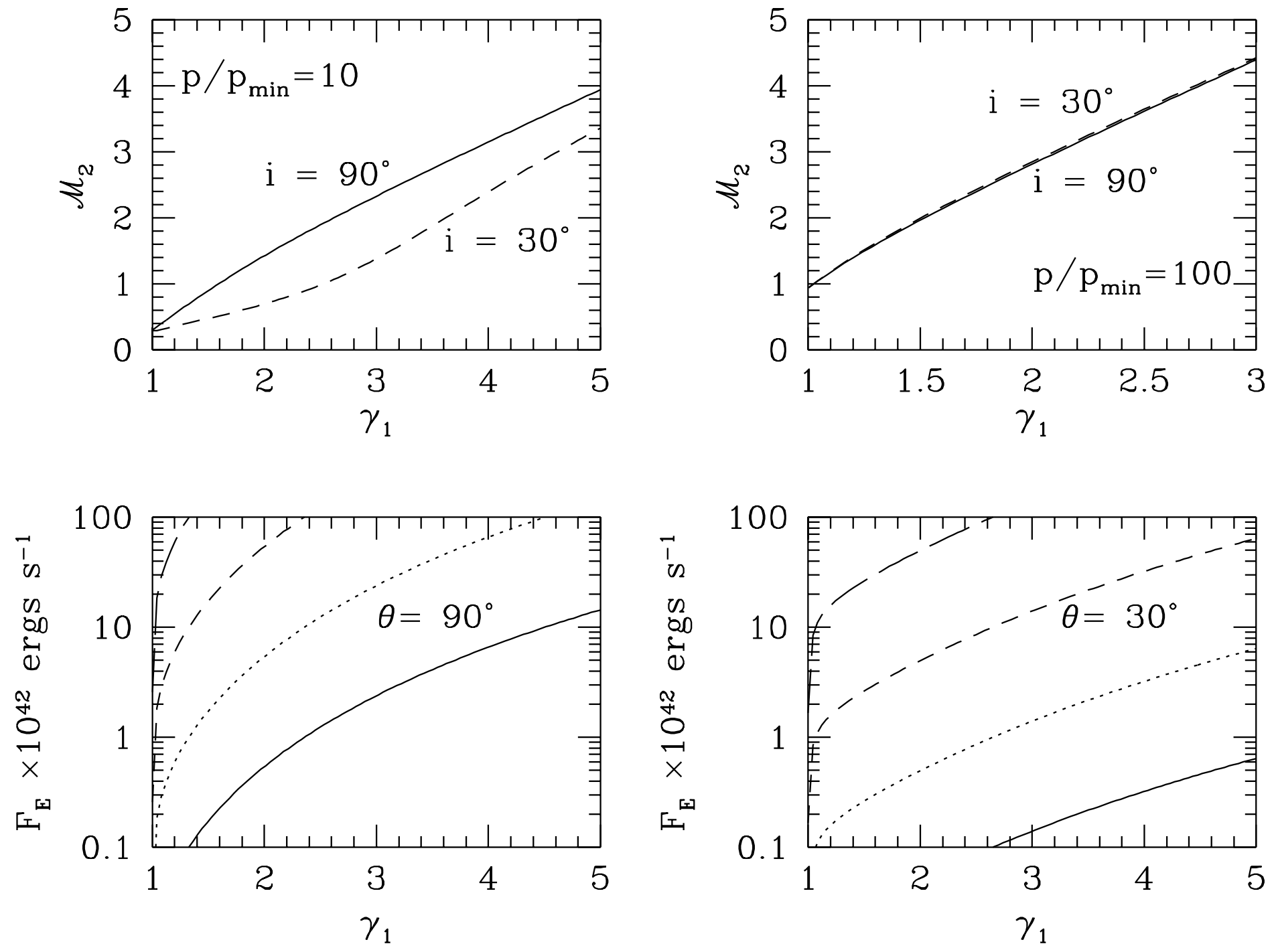

FIGURE II *Upper panels:* Mach Number at $\Theta = 17''$ as a function of pc-scale Lorentz factor and jet inclination. *Lower panels:* Jet Energy Flux for values of $p/p_{\rm min} = 0.1, 1, 10$ and 100(increasing from lower to upper curves).

Similar considerations also apply to NGC 6251 (see Bicknell, 1993). In that case the energy flux is probably better estimated from reliable spectral index variations.

Modeling of the large scale jet in NGC 315 (Bicknell, these proceedings) predicts a velocity decline between $\Theta = 17''$ and $\Theta = 100''$ by a factor of a few. If the jet velocity is approximately 0.6 c at $\Theta = 17$, this decline suffices to account for the equality of surface brightnesses at $\Theta \approx 100''$.

## KNOTS AND SIDEDNESS

The small number of FRI sources that have been imaged with VLBI exhibit *sub*luminal motion (e.g. M87, Reid *et al.*(1989); NGC 315, Venturi *et al.*(1993)) giving one cause to question the validity of relativistic velocities on small scales in FRI sources. The physical case is neatly summarized in figure III where the brightness ratio of two oppositely directed jets is plotted against the apparent $\beta$. Obviously, a high brightness ratio and a low transverse velocity ($R > 50$ and $\beta < 0.5$ for the pc-scale jet in NGC 315) imply a prohibitively small angle to the line of sight for a source with already a large projected size on the sky. A similar angle is inferred for M87. Lind and Blandford (1985) emphasized that knots are most naturally interpreted as shocks. Given this, and the models of the flux variations of BL-Lac in which the knots are reverse shocks caused by variations in the flow velocity (Hughes, Aller and Aller, 1989a,b) it is natural to consider the implications of reverse shocks in FRI jets. Even a weak shock can move rapidly backwards against the jet flow (with a relative velocity of $c/\sqrt{3}$) leading to a low velocity in the observer's frame. This situation is summarized in the right panels of figure III where the observed pre- and post-shock fluid velocities are plotted for different observed shock velocities and shock angles as a function of the shock pressure ratio. Clearly, both velocities are considerably larger than the observed shock velocity especially for oblique shocks where the parallel component of the fluid velocity is unaffected. The limiting values for the post-shock velocity correspond to the cases of weak and strong shocks respectively.

## THE EMISSION LINE FILAMENTS IN CENTAURUS A

As indicated in the introduction the extranuclear emission lines of radio galaxies pose interesting problems relating to their excitation and temperature. Although the emission line filaments in Centaurus A are of low luminosity ($\sim 10^{38}\mathrm{ergs\,s^{-1}}$) compared to emission line regions in powerful FRII radio galaxies, the proximity of Centaurus A means that detailed study is extremely rewarding. Morganti *et al.*(1991) have obtained excellent [OIII] images together with high and low dispersion spectra of the "inner" and "outer" filaments of Cen A showing the presence of both high and low excitation lines and large turbulent velocities ($\sim 200 - 400\,\mathrm{km\,s^{-1}}$). Both the inner and outer filaments show the apparent effects of disruption by the radio plasma in Centaurus A.

Morganti *et al.*have argued that the filaments are excited by a narrow beam of radiation from the core of the galaxy, showing that if this is the case, then the beaming cannot be due to obscuration by a dusty torus and must be intrinsic. My colleagues (Ralph Sutherland, Mike Dopita, David Singleton) and I, in a series of papers (Bicknell, 1991, Sutherland, Bicknell and Dopita, 1993 (SBD), Bicknell, Dopita and Singleton 1993) have developed a model for these filaments which invokes interaction between a radio jet and dense clouds along its path. Low dynamic range images (Burns 1993) of the *Middle Lobe* region of Centaurus A show that the *Outer Filaments* are coincident with the southern region of the middle lobe but that there is no obvious radio emission coincident with the *Inner Filaments*. However, the inner filaments lie along the extrapolated path of the inner jet. Clearly higher dynamic range AT or VLA images are required in view of the fact that our model predicts that radio plasma is still flowing in the region of the inner filaments.

Our model involves: (1) Disruption of the molecular cloud by the Kelvin-Helmholtz instability acting at the jet-cloud interface (2) Production of super-sonic turbulent velocities in the molecular material (3) Dissipation of the turbulence via radiative *autoionizing* shocks and (4) Simultaneous production of high and low excitation lines through density and/or geometric effects. The idea of autoionizing shocks was first introduced by Binette, Dopita and Tuohy (1985). Let us now consider the key elements of this model.

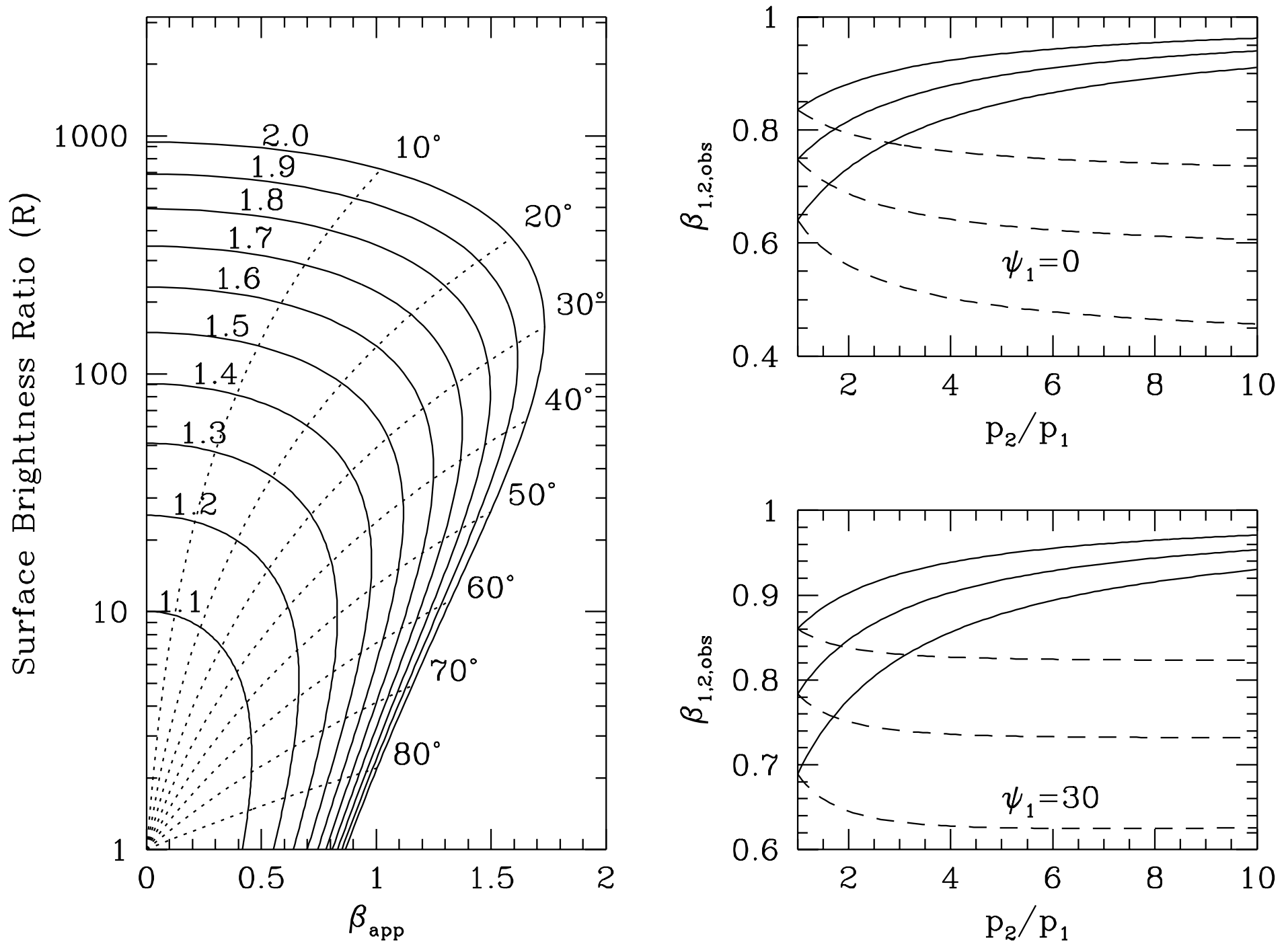

FIGURE III *Left Panel:* The surface brightness ratio of two equal, oppositely directed jets as a function of the apparent jet velocity for a range of Lorentz factors between 1 and 2. *Right Panels:* The pre- and post-shock velocities corresponding to shock velocities of 0.1, 0.3 and 0.5$c$ as a function of shock pressure ratio both for a normal and an oblique shock ($\psi_1$ is the angle between the shock normal and the pre-shock velocity in the shock frame.)

**The Kelvin-Helmholtz Instability**

There is a well known limit on the Mach number for the K-H instability to occur, namely, $\mathcal{M}\, cos\phi < \left(1 + (c_{s,2}/c_{s,1})^{2/3}\right)^{3/2}$ where $\phi$ is the angle between the perturbation and the flow, $c_s$ is the sound speed and subscripts 1 and 2 refer to the jet and cloud respectively (Miles 1958, Fejer and Miles 1963). Strictly, this inequality refers to the case of equal specific heat ratios; nevertheless it should be generally indicative. In the case in point, the jet Mach number clearly should not be much larger than unity in order for the instability to be effective, consistent with the idea that the jets in the outer parts of FRI sources are transonic.

### Simulations of Cloud Ablation

In view of the potential relevance of the Kelvin-Helmholtz instability, simulations for flows with Mach numbers $\lesssim 1$ and a high density ratio are useful to ascertain whether the physical reason for some of the features of the filaments can be identified. Simulations with a density ratio $\sim 10^{-4}$ are difficult to perform because of the lower growth rates and the Courant criterion. However, simulations for a density ratio of 100 are relatively straightforward. Figure IV shows the evolution of the density in such a simulation for $\mathcal{M} = 0.5$. The jet is the upper gray region and the cloud is the lower dark region. The jet initially makes an angle of 5° with the cloud and this is the reason for the intermediate colored wedge of material at the left. This simulation exhibits many of the observed features of the filaments, in particular the wave-like region of knot A, B and C with its long drawn out filament, the filamentary structure of the surface waves and a socket-like feature at the end of the long filament in frame 7 which appears and disappears as the filament waves in and out of regions of differing velocity. Further downstream, arcs of gas which have been totally ablated from the main cloud. Another welcome feature of these simulations are the sound waves and weak shocks which are driven into the dense material by the turbulence at the interface. Some of these arise from the collision of filaments on the interface. These filaments have velocities of order 10 – 20% of the free stream velocity. The density ratio in this simulation is large, but not large enough to show the development of highly supersonic turbulence since the sound speed in the dense region is not low enough. However, given that the Kelvin-Helmholtz instability is capable of producing velocities in the dense material equal to a substantial fraction of the free-stream velocity, then as the density ratio increases, the turbulent velocities should become more supersonic. This point is currently being checked with higher density ratio simulations.

### The Energy Budget

The jet energy flux incident on a filament of cross-sectional area $A$ is:

$$F_E = 1.0 \times 10^{37} \frac{\gamma_{\rm jet}^{3/2}}{\gamma_{\rm jet}-1} \eta^{-1/2} \left(\frac{n_{\rm ism}}{10^{-3}}\right) \left(\frac{T_{\rm ism}}{10^7}\right)^{3/2} \left(\frac{A}{10^4 {\rm pc}^2}\right) \mathcal{M}_{\rm jet} \left(1 + \frac{\gamma_{\rm jet}-1}{2}\mathcal{M}_{\rm jet}^2\right) {\rm ergs\ s^{-1}} \qquad (6)$$

For $A \sim 4 \times 10^4$ pc, $\eta \approx 0.1$ and $\mathcal{M} \sim 1$ the incident power is $\approx 2 \times 10^{39}\,{\rm ergs\ s^{-1}}$ adequate to account for the $H\beta$ emission-line luminosity $\sim 10^{37}\,{\rm ergs\,s^{-1}}$ assuming that most of the incident power is dissipated in radiative shocks and allowing for a factor $\sim 100$ between the total emission line luminosity and the $H\beta$ luminosity.

### Order of magnitude estimates for the Balmer Line Luminosity

Given that the turbulent supersonic velocity field produced by the Kelvin-Helmholtz instability is dissipated through the occurrence of strong shocks, it is possible to estimate the Balmer line luminosity from the data on the turbulent velocity field. The geometry and different ionization zones of cloud-cloud collisions is described in detail in Sutherland, Bicknell and Dopita (1993).

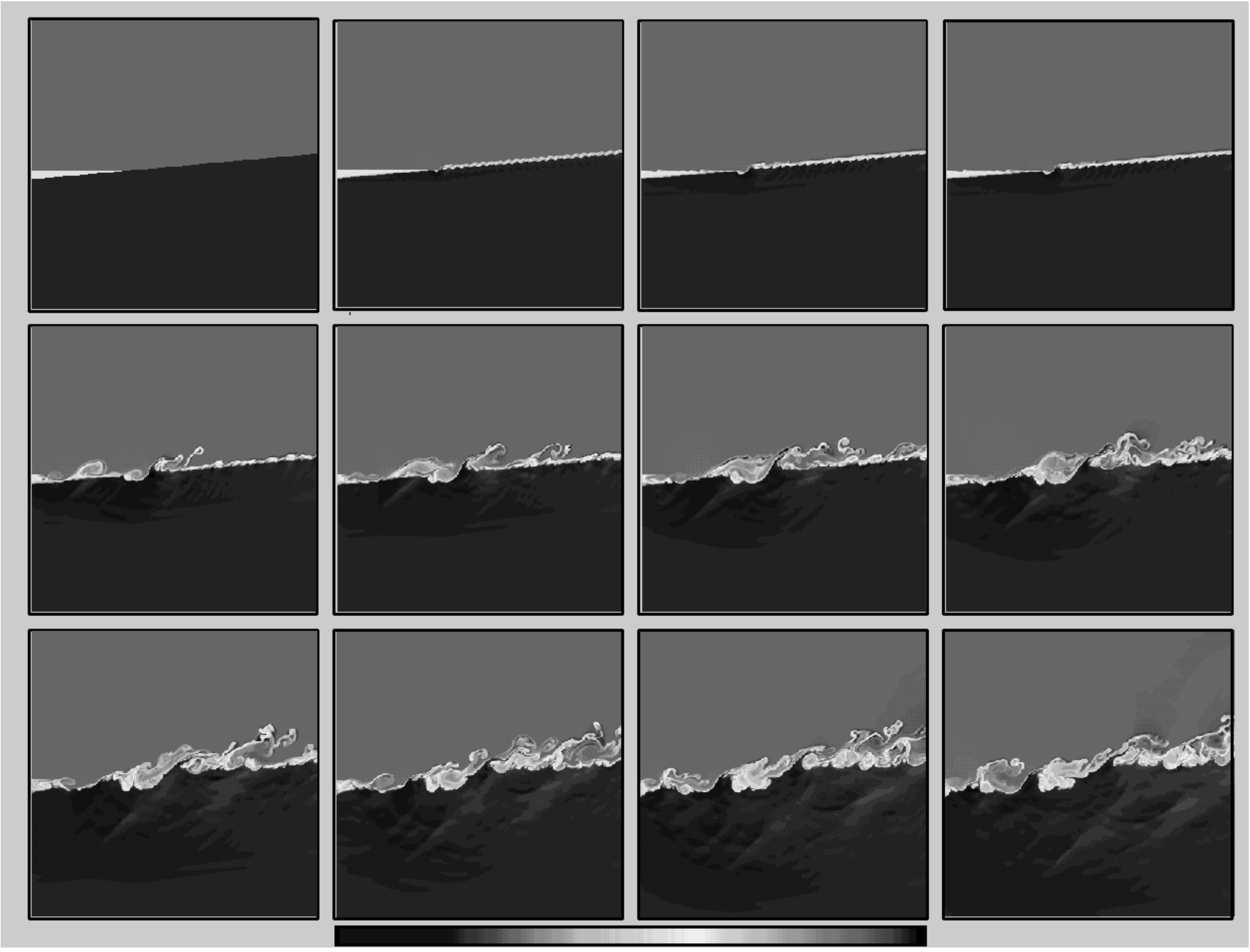

FIGURE IV The evolution of the density (log scale) in a 2D simulation of the Kelvin-Helmoltz instability. Note the production of filaments and ablated clouds at the interface.

When two equal density clouds collide with a velocity $\Delta v \approx 200\,\mathrm{km\,s^{-1}}$ the temperature produced behind each of the strong shocks propagated into the clouds is:

$$T_{\rm sh} \approx \frac{3}{64}\frac{\mu\, m_p}{k}(\Delta v)^2 = 1.3\times 10^5 \left(\frac{\Delta v}{200\,\mathrm{km\ s^{-1}}}\right)^2 \qquad (7)$$

and this can be high enough to produce an ionizing flux of UV and soft X-ray radiation. The cooling time behind the shocks is short so that all the internal energy is radiated into the upstream and downstream plasma. Thus, the $H\beta$ luminosity (for optically thick photoionization) is given by:

$$L_{H_\beta} = 1.5\times 10^{36}\,\mathrm{ergs\ s^{-1}}\, f\, n_0 \left(\frac{\Delta v}{200\mathrm{km\ s^{-1}}}\right)^3 \left(\frac{A_{\rm sh}}{10^4\mathrm{pc}^2}\right) \qquad (8)$$

where $f = \left[\int_{\nu_0}^{\infty} (\nu_0/\nu)\Lambda_\nu\, d\nu\right]\left[\int_0^\infty \Lambda_\nu\, d\nu\right]^{-1}$ parametrizes the ratio of the flux of ionizing photons to post-shock luminosity and is approximately 0.8 for non-equilibrium cooling as described in SBD.

When cross-sections of elements other than Hydrogen are taken into account, model calculations show that it the exponent of $\Delta v$ is closer to 2.4 (see Dopita, these proceedings). Nevertheless, equation 8 is quite accurate in the velocity range of interest and shows that the luminosity of the knots is comfort-ably accounted for by the combination of the velocity field observations and this simple physical picture.

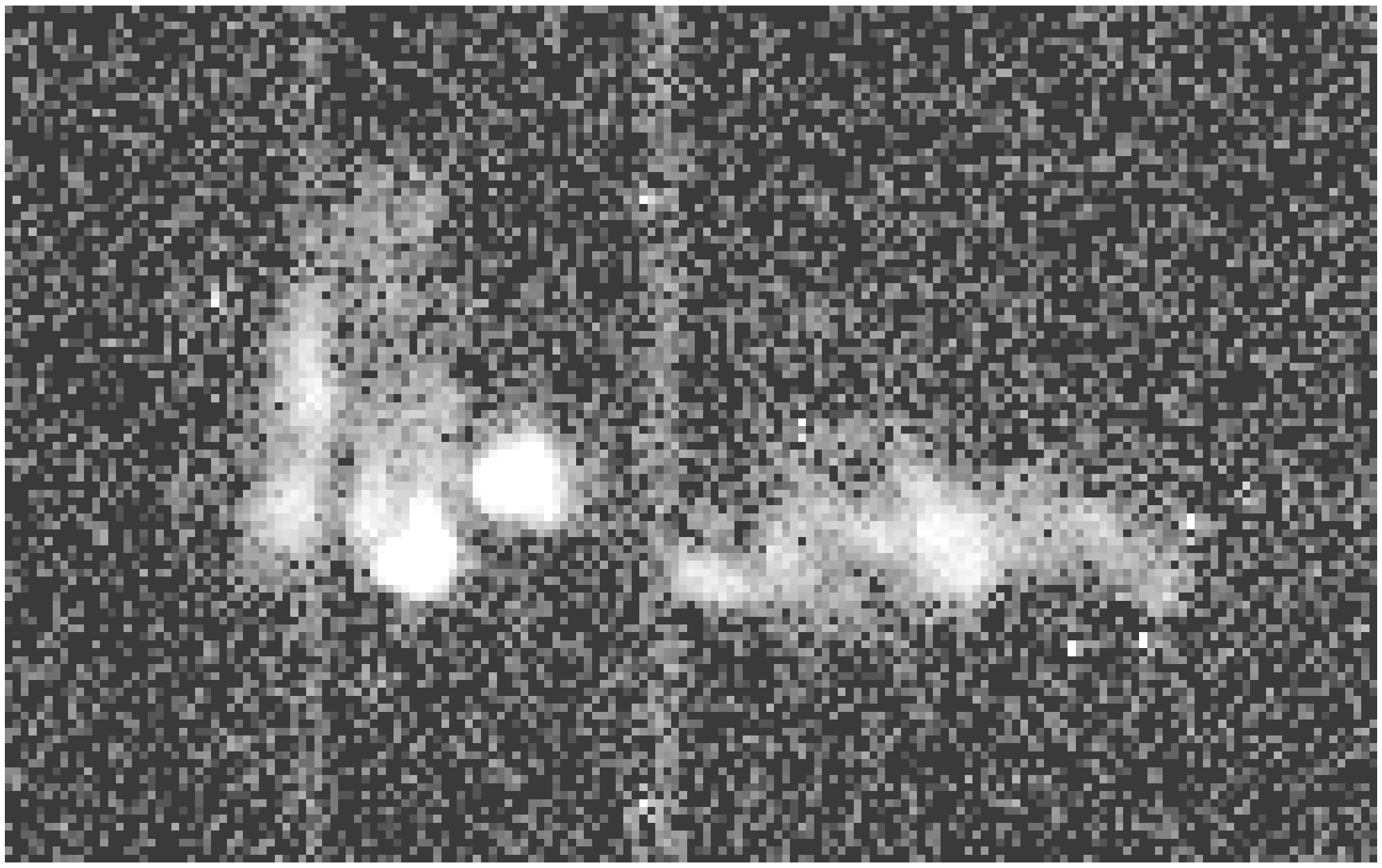

FIGURE V A spatial-velocity image of the inner filaments based on [OIII]$\lambda$5007. The horizontal axis is the spatial distance along the slit. The vertical axis is velocity. The direction of the nucleus is to the left.(From Bicknell, Dopita and Singleton, 1993)

The modeling carried out in SBD also showed that both high excitation and low excitation lines could be produced if different regions of the interface have different density. These models provided a good fit to the observed spectrum including the temprature sensitive line $[OIII]\lambda 4363$ which is produced in the high temperature region behind a shock.

## RECENT AAT OBSERVATIONS OF CENTAURUS A

Figure V shows an AAT spectrum in the vicinity of [OIII]$\lambda$5007 with approximately $30\,\mathrm{km\,s^{-1}}$ resolution. The slit passes through the middle of the inner filaments and includes knots A, B and C. This spectrum shows some dramatic features including large regional differences in the velocity field, evidence for entrainment in the form of features emanating from the brightest knots and extending in the velocity direction by about $700\,\mathrm{km\,s^{-1}}$ and evidence for a surface-wave in the right hand part of the image. confirming the physical picture presented above.

## CONCLUSIONS

That beaming is important in the determining the radio characteristics of Radio Galaxies seems to be unavoidable. The overwhelming impression that one gains from working on the VLA and VLBI data of FRI radio galaxies is that the notions of relativistic jet asymmetries, jet deceleration and FRI-BL-Lac unification are all consistent. The case for beamed optical radiation is not so clear cut. Nothwithstanding the popularity of beamed ionizing radiation, I have presented arguments that the emission line filaments in the nearest Active Galaxy, Centaurus A, are excited by their interaction with an, as yet undetected, radio jet. Perhaps one of the most appealing aspects of this model is its high degree of internal consistency: The morphology, internal velocities and spectra of the filaments are all explained in terms of the interaction between a $\sim 1500 - 50000\,\mathrm{km\,s^{-1}}$ radio jet and dense molecular clouds along its path. The model makes a clear prediction for the flux of the ultraviolet line CIV$\lambda$1549 (see SBD). Clearly, this model, if substantially correct, has ramifications to other emission line regions (see Koekemoer, Bicknell and Dopita, these proceedings). The bottom line in all of these calculations is the unification of the optical line emission and the local kinematics.